\documentclass[10pt,conference]{IEEEtran}
\usepackage[left=0.7in, right=0.73in, top=0.7in, bottom = 0.98in]{geometry}
\usepackage{algorithm2e}
 \RestyleAlgo{ruled}
\usepackage{graphicx}
\usepackage{setspace}
\usepackage{adjustbox} 
\DeclareGraphicsExtensions{.pdf,.jpeg,.png}
\usepackage{enumitem}
\usepackage[noadjust]{cite}

\usepackage{amsmath}
\usepackage{lipsum}
\usepackage{mathtools}
\usepackage{cuted}
\usepackage{array}
\usepackage{subfigure}
\usepackage{amssymb}
\usepackage{caption}
\usepackage{multirow}
\usepackage{makecell}

\usepackage{xcolor}
\def\BibTeX{{\rm B\kern-.05em{\sc i\kern-.025em b}\kern-.08em
ThedeltaT\kern-.1667em\lower.7ex\hbox{E}\kern-.125emX}}

\begin{document}
\SetKwComment{Comment}{/* }{ */}

\title{Joint Uplink and Downlink Resource Allocation and Antenna Activation for Pinching Antenna Systems}

\author{Shreya Khisa, Ali Amhaz, Mohamed Elhattab, Chadi Assi, Sanaa Sharafeddine 
}
\maketitle

\begin{abstract} 
In this paper, we explore a novel joint uplink and downlink framework utilizing a pinching antenna system (PASS). We consider two waveguides, one dedicated to transmission and one to reception, and both of them are connected to a base station (BS). Each type of waveguide consists of several pinching antennas (PAs) in some preconfigured positions. In this framework, we assume the BS can serve downlink and uplink user equipments (UEs) at the same time using the same spectrum resources through the presented PASS. In this aspect, we formulate a sum rate optimization problem that jointly optimizes the antenna activation factor, the BS transmit power, and the UE's transmit power, subject to power budget constraints for the BS and the UEs, as well as minimum rate requirements for the UEs. The formulated problem is highly non-convex and difficult to solve directly. Hence, we divide the main problem into two sub-problems: the antenna activation sub-problem and the power allocation sub-problem. Then, we solve the antenna activation problem utilizing a distance and spatial correlation-based algorithm. Meanwhile, the resource allocation problem is solved using a successive convex approximation (SCA)-based algorithm. Numerical results show that our proposed framework can achieve around 60-90\% performance gains over its time division duplex (TDD) where the uplink and downlink transmissions are served in different orthogonal time slots.
\end{abstract}

\section{Introduction}
The next generation of wireless communication (6G) is expected to bring a major technological transformation that has not been seen in previous generations of communication systems \cite{liu2025pinching}. The unprecedented surge in applications and services, mainly driven by artificial intelligence and automation, will require increased data rate, ultra-low latency, high reliability, and high spectral efficiency. One notable technology that has evolved over the years and has shown tremendous performance in terms of enhancing spectral efficiency is the multiple-input and multiple-output (MIMO) system \cite{liu2025pinching}. Looking back at the evolution from 2G to 5G, MIMO has consistently advanced wireless communication networks by delivering significant improvements in multiplexing and diversity gains. Since then, MIMO has evolved in more advanced forms such as massive MIMO, gigantic MIMO, and recently continuous aperture arrays (CAPA) \cite{liu2025pinching}. Although these technologies can improve spectral efficiency, they come with higher costs, such as high computational complexity, heavy channel estimation overhead, and increased implementation cost \cite{liu2025pinching}. 
\par In light of these challenges, flexible antenna systems such as pinching antenna systems (PASS) (and others such as reconfigurable intelligent surface and movable antenna) have recently gained significant attention from researchers and academic communities. PASS was first introduced by NTT DOCOMO at the Mobile World Congress (MWC) in 2021 \cite{suzuki2022pinching}. The architecture of PASS consists of two essential parts: dielectric waveguides and separate dielectric pinching antennas (PAs). The purpose of waveguides is to serve as the transmission media that can carry signals over long distances with low attenuation. Unlike conventional MIMO, which suffers from large-scale fading due to the non-line-of-sight (NLoS) communication, PASS signals passed through the waveguide can be radiated to free space at each PA's location. In particular, PAs are implemented by employing small dielectric particles to excite particular points along a waveguide \cite{yang2025pinching}. As a result, the PA functions as a leaky wave antenna, generating well-controlled radiation spots. Hence, utilizing PASS, direct LoS communication can be possible, which increases the spectral efficiency. 
\par Recently, several works have been proposed considering different scenarios of PASS. For example, the authors in \cite{wang2025sum} presented a downlink sum rate maximization problem with an antenna activation algorithm for pinching antenna and rate-splitting multiple access (RSMA). Meanwhile, the authors in \cite{10912473} studied the antenna activation problem of PAs considering non-orthogonal multiple access (NOMA) in the downlink. In \cite{10896748}, a downlink rate maximization problem has been proposed with continuous PA location activation. On the other hand, uplink performance analysis for PASS is presented in \cite{hou2025performance}. The authors in \cite{10909665} investigated the user fairness problem for uplink PASS with continuous PA location activation. In \cite{bereyhi2025mimo}, explored the potential of deploying PASS for uplink and downlink transmission in multiuser MIMO settings. In particular, they proposed a time-division duplex (TDD) mode-based system model where the same waveguide was used for both downlink communication and uplink communication, but in different time slots. They proposed a rate maximization problem with the consideration of continuous PA location activation in both uplink and downlink. 
\par Even though there are several works of rate maximization optimization for individual uplink, downlink, and TDD mode, to the best of our knowledge, there has been no work that investigates PASS systems that use the same time frequency resource for both uplink and downlink communications. Hence, in this paper, we propose a novel joint uplink and downlink communication model for PASS, considering one transmitting waveguide and one receiving waveguide and multiple uplink and downlink user equipments (UEs). We assume that PAs can be activated in the preconfigured positions. Particularly, we analyze two scenarios: in the first scenario, we assume that there exists an inter-waveguide interference between the transmitting waveguide and the receiving waveguide. In the second scenario, we assume that BS can perfectly estimate the channel between each PA at the transmitting waveguide and each PA in the receiving waveguide. Hence, there exists no inter-waveguide interference between them. In this aspect, we solve an optimization problem where we jointly optimize BS and UE transmission power and antenna activation factor with the objective to maximize the achievable sum rate. The formulated optimization problem is highly non-convex. We adopt a distance- and spatial-correlation-based approach for antenna activation, and the resource allocation problem is solved using a successive convex approximation (SCA)-based approach. Simulation results demonstrate that our proposed scheme can achieve around 60-90\% performance gains over the TDD-based approach.
\section{System Model}
\begin{figure}
    \centering
\includegraphics[width=1\columnwidth]{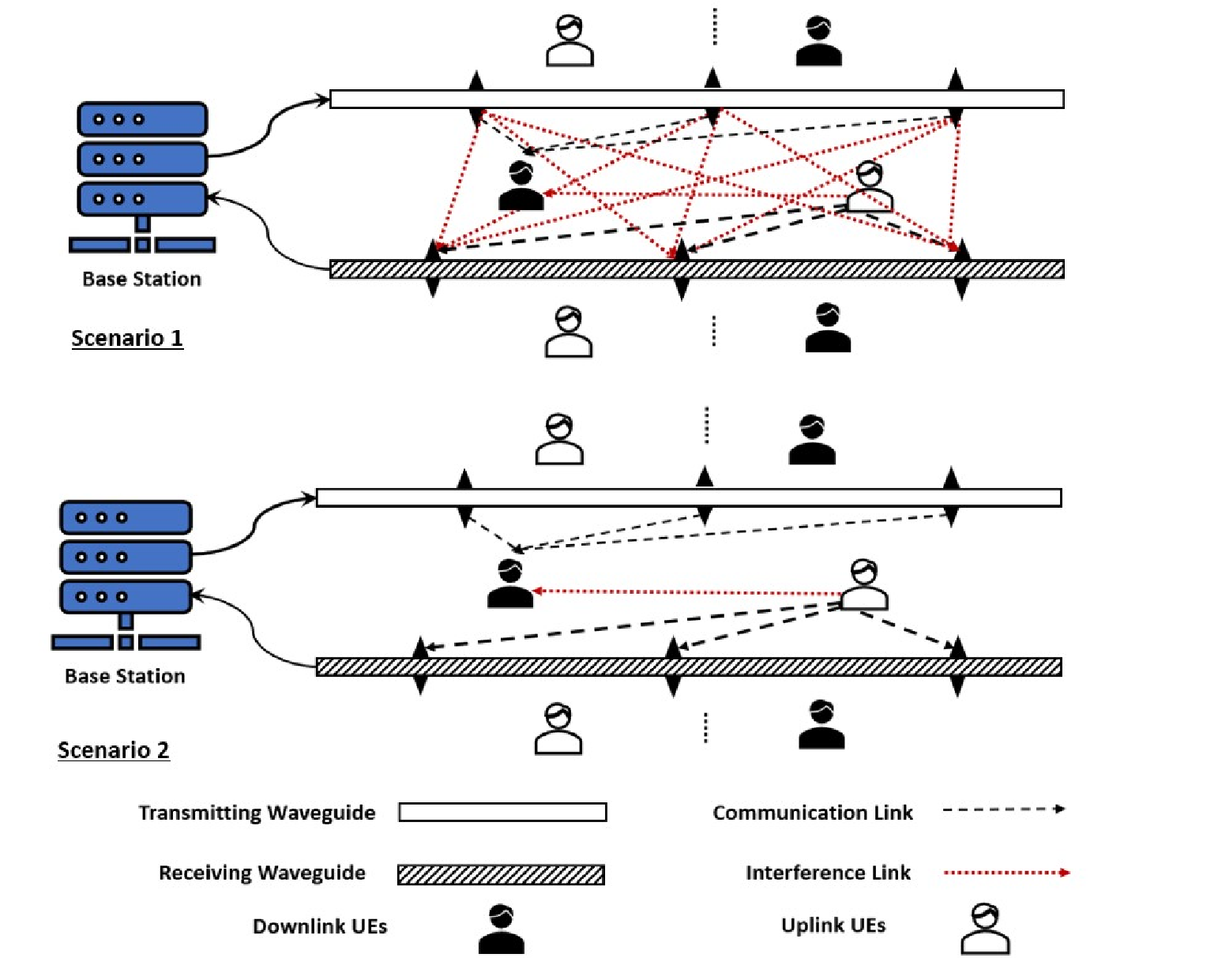}
    \caption{PASS-based Transmission Framework}
    \label{fig1}
\end{figure}

As shown in Fig. \ref{fig1}, we assume a base station (BS), which is equipped with a transmitting waveguide $t$ and a receiving waveguide $r$, each of which is fed by an RF chain and comprises $N$ PAs. We assume that $N$ PAs are pre-configured in their designated positions, and they can be activated and deactivated as needed. Without loss of generality, it is assumed that the waveguides are deployed parallel to the $x$-axis at a height $h$, as shown in Fig.\ref{fig1}. The $y$-axis coordinate of the waveguide $t$ is denoted by , $y_t$ and it spans from its feed point $lt_0=[0,y_t,h]^T$ to $[D,y_t,h]^T$ with a length of $D$. The coordinate of the $n$-th PA on the waveguide $t$ is given by $lt_{n}=[x_{n},y_{t},h]^T$ where $x_{n}$ denotes the $x$-axis coordinate of the $n$-th PA on the waveguide $t$. The waveguide $t$ serves $K$ downlink users, and the coordinates of the $k$-th downlink user are located at $u_k=[x_k,y_k,0]^T$. Accordingly, we can model $y$-axis coordinate of the waveguide $r$ as $y_r$ and it spans from its feed point $lr_0=[0,y_r,h]^T$ to $[D,y_r,h]^T$ with a length of $D$. The coordinate of the $n$-th PA on the waveguide $r$ is given by $lr_{n}=[x_{n},y_{r},h]^T$ where $x_{n}$ denotes the $x$-axis coordinate of the $n$-th PA on the waveguide $r$. The waveguide $r$ serves $U$ uplink users, and the coordinates of the $u$-th uplink user are located at $u_u=[x_u,y_u,0]^T$. We can define the set for downlink UEs as $\mathcal{K}=[1,\dots,K]$ and the set for uplink UEs as $\mathcal{U}=[1,\dots,U]$. As depicted in fig. \ref{fig1}, we consider two scenarios for our proposed framework:
\begin{itemize}
    \item \textbf{Scenario 1}: In the first scenario, we assume that there exists an interference between the transmitting waveguide and receiving waveguide due to the uplink and downlink communication happening using the same time and frequency resource.
    \item \textbf{Scenario 2}: In the second scenario, we assume that there exists no interference between the transmitting and receiving waveguides due to the fact that both are connected to the BS, and PAs are installed in preconfigured positions. Hence, BS fully estimates the channel state information (CSI) between the transmitting PAs and receiving PAs.
\end{itemize}

\section{Transmission model}
\subsection{Downlink Channel Modeling}
The free space channel between the antennas of the waveguide $t$ and the $k$-th user is given by \cite{10912473},
\begin{equation}
    {h}_{n,k}=\frac{{\eta} e^{-j\frac{2\pi}{\lambda}\|lt_{n}-u_k\|}}{\|lt_n-u_k\|},
    \label{e1}
\end{equation}
where $\eta=\frac{c}{4\pi f_c}$, $c$ is the speed of light, $f_c$ represents the carrier frequency, and $\lambda$ is the wavelength. $\|lt_n-u_k\|$ represents the Euclidean distance between UE-$k$ and $n$-th PA in waveguide $t$.
With the assumption of lossless in-waveguide propagation, the in-waveguide channel vector can be given by \cite{10912473},
\begin{equation}
    {{g}}_{n}^t=e^{-j\frac{2\pi}{\lambda_{g_t}}\|{lt}_{n}-{lt}_{0}\|},\label{e2}
\end{equation}
where $\lambda_{g_{t}}=\frac{\lambda}{\eta_{eff}}$ represents the waveguide wavelength and $\eta_{eff}$ represents the effective refractive index of the waveguide. $\|{lt}_{n}-{lt}_{0}\|$ represents the Euclidean distance between $n$-th PA to the feed-point of transmitting waveguide.
Thus, the channel between user $k$ and the $n$-th PA can be given as,
\begin{equation}
    \textbf{h}_{k}=\left[\delta_1h_{1,k}g_{1}^t, \dots, \delta_Nh_{N,k}g_{N}^t\right]^T, \label{e3}
\end{equation}
where $\delta_{n} \in \{0,1\}$ where $\delta_n = 0$ indicates the $n$-th PA on the waveguide $t$ is deactivated and $\delta_n = 1$ denotes PA is activated. 
Using \eqref{e1}, \eqref{e2} and \eqref{e3}, we can model the channel between user $k$ and the waveguide $t$ as follows
\begin{equation}
  {h}_k=\sum_{n=1}^N\delta_n \frac{{\eta} e^{-j 2 \pi{\left(1/\lambda||lt_n-u_k||+1/\lambda_{g_t}||lt_0-lt_n||\right)}}}{||lt_n-u_k||}. \label{e4}
\end{equation}
\subsection{Downlink Transmission Model}
Now, the transmitted signal from BS to the $K$ downlink users is given as,
\begin{equation}
\hat{s}_t=\sum_{k=1}^K\sqrt{p_{k}}s_k,
\end{equation}
where $s_k$ represents the signal for UE-$k$ and $p_{k}$ represents the transmit power for $s_k$. 
The signal received from the BS at the $k$-th UE in the downlink is given by,
\begin{equation}
y_k={h}_{k}\sqrt{\frac{1}{\sum_{n=1}^N \delta_n}}\hat{s}_t+\sum_{u\in U}h_{u,k}x_u+n_k,
\end{equation}
where
$n_k \in \mathcal{CN}(0, \sigma_k^2)$ denotes the additive white Gaussian noise (AWGN) at the $k$-th user. $\sum_{u\in U}h_{u,k}x_u$ represents the channel between $u$-th UE and $k$-th UE. This channel accounts for both large-scale and small-scale path loss effects.
In this work, we assume that the transmit power of activated PAs is equally allocated \cite{wang2025modeling}. Hence, 
$\sqrt{1/\sum_{n=1}^{N}\delta_n}$ represents the power allocation coefficient for activated PAs in each waveguide.
Therefore, the achievable rate analysis at UE-$k$ is presented in \eqref{Rk}.
\begin{table*}[!ht]
\begin{equation}
R_{k}= 
\log_2\left(1+\frac{|{h}_k|^2 \frac{p_k}{\sum_{n=1}^N \delta_n}}{\sum_{k', k' \neq k}^K|{h}_k|^2\frac{p_k'}{\sum_{n=1}^N \delta_n}+\sum_{u \in U}|h_{u,k}|^2P_u+\sigma_k^2}\right).\label{Rk}
\end{equation}
\end{table*}
\subsection{Uplink Channel Modeling}
Channel coefficient between UE-$u$ and PA-$n$ on the waveguide $r$ can be given by \cite{bereyhi2025mimo},
   \begin{equation}
       h_{n,u}=\frac{\eta e^{-j\frac{2\pi}{\lambda}\|{lr}_{n}-{u}_u\|}}{\|{lr}_{n}-{u}_{u}\|},
   \end{equation}
where $\|lr_n-u_u\|$ represents the Euclidean distance between UE-$u$ and $n$-th PA in waveguide $r$.
We can model the in-waveguide propagation as follows \cite{bereyhi2025mimo},
\begin{equation}
    {g}_{n}^r=e^{-j\frac{2\pi}{\lambda_g}\|{lr}_{n}-{lr}_{0}\|},
\end{equation}
where $\|{lr}_{n}-{lr}_{0}\|$ represents the Euclidean distance between $n$-th PA and receiving waveguide feedpoint.
Following \eqref{e3}, we model the uplink channel as follows, 
   \begin{equation}
    \textbf{h}_{u}=\left[\beta_1h_{u,1}g_{1}^r, \dots, \beta_Nh_{u,N}g_{N}^r\right]^T,
\end{equation}
where $\beta_{n} \in \{0,1\}$ where $\beta_n = 0$ indicates the $n$-th PA on the waveguide $r$ is deactivated and $\beta_n = 1$ denotes PA is activated. 
Following \eqref{e4}, we can model the uplink channel as,
\begin{equation}
    {h}_u=\sum_{n=1}^N\beta_n\frac{{\eta} e^{-j 2 \pi\left(1/\lambda||lr_n-u_u||+1/\lambda_{g_r}||lr_0-lr_n||\right)}}{||lr_n-u_u||}.
\end{equation}
\subsection{Uplink Transmission Model}
The signal transmitted by UE-$u$ is given by
\begin{equation}
    x_u=\sqrt{P_u}s_u.
\end{equation}
   We assume that both uplink and downlink transmissions occur at the same time frequency resource. Hence, we can model the channel between the transmitting waveguide and the receiving waveguide as follows,
\begin{equation}
    \textbf{H}[TR]=\sqrt{PL({d}})\left(\sqrt{\left(\frac{\lambda_x}{1+\lambda_x}\right)}m_x+\sqrt{\left(\frac{1}{1+\lambda_x}\right)}\hat{m}_x\right),
\end{equation}
where $\textbf{H}[TR] \in \mathbb{C}^{N \times N}$, $m_x$ represents the LoS component, $\hat{m}_x$ represents the NLoS component, $\lambda_x$ represents
the Rician factor and $PL(d)$ denotes the large-scale path loss between two waveguides. The received signal at the BS from $u$-th UE can be given as,
\begin{align}
    y_r^u=\frac{1}{{\sum_{n=1}^N \beta_n}}(h_ux_u+\hat{\textbf{g}}_r \textbf{H}[TR]\hat{\textbf{g}}_t^H\hat{s}_t+n_u),
\end{align}
where the $\hat{\textbf{g}}_r \textbf{H}[TR]\hat{\textbf{g}}_t^H\hat{s}_t$ represents the inter-waveguide interference (Scenario 1), $n_u$ represents the AWGN with zero mean and $\sigma^2$ variance. Since the transmit signal from the activated PAs is received only by the activated PAs on the receiving waveguide, the in-waveguide transmission for the transmitting and receiving waveguides can be expressed as follows:
\begin{equation}
    \hat{\textbf{g}}_r=\left[\beta_1g_{r,1}, \dots, \beta_Ng_{r,N}\right] \in \mathbb{C}^{1 \times N},
\end{equation}
\begin{equation}
    \hat{\textbf{g}}_t=\left[\delta_1g_{t,1}, \dots, \delta_Ng_{t,N}\right] \in \mathbb{C}^{1 \times N}.
\end{equation}
The achievable rate to decode the signal of UE-$u$ at the BS can therefore be derived as \eqref{Ru}.
\begin{table*}[!t]
\begin{equation}
    R_u=
\log_2\left(1+\frac{\frac{P_u}{\sum_{n=1}^N{\beta_n}}|{h}_u|^2}{\sum_{u', u' \neq u}^U\frac{P_{u'}}{\sum_{n=1}^N {\beta_n}}|{h}_{u'}|^2+\frac{1}{\sum_{n=1}^N {\beta_n}}|\hat{\textbf{g}}_r\textbf{H}[TR](\sum_{k \in K}p_k)\hat{\textbf{g}}_t^H|^2+\sigma^2/\sum_{n=1}^N{\beta_n}}\right). \label{Ru}
\end{equation}
\end{table*}
It should be noted that for the second scenario, the term "inter-waveguide interference" is assumed to be negligible; that is, because in scenario 2, we assume that we utilize pre-configured positions of PA. Hence, BS can perfectly know the CSI among the PAs situated in transmitting and receiving waveguides \cite{guo2024movable}. Hence, it can fully cancel out the interference among the waveguides. Therefore, no inter-waveguide interference is considered.
\section{Problem Formulation}
We aim to maximize the sum rate of both downlink and uplink UEs while guaranteeing the minimum rate requirements by jointly optimizing the antenna activation factor, BS transmit power, and uplink UEs transmit power. The formulated optimization problem can be defined as follows: 
\allowdisplaybreaks
\begin{subequations}
\label{prob:P1}
\begin{flalign}
\centering
 &\mathcal{P}_1: \max_{\substack{\textbf{P}_K, \, \textbf{P}_U,\, \boldsymbol{\beta}, \boldsymbol{\delta}}} \quad \quad \sum_{k \in \mathcal{K}} R_{k}+\sum_{u \in \mathcal{U}} R_u,\:\label{const1} \\
 &\text{s.t.} \quad \sum_{k=1}^Kp_k \le P_t,\label{c10}\\
 & \qquad P_{u} \le P_u^{max}, P_{u} \ge 0, \forall u \in \mathcal{U},\label{c1}\\
 & \qquad R_k \ge R_{th,k},  R_u \ge R_{th,u}, \forall k \in \mathcal{K}, \forall u \in \mathcal{U}, \label{c3}\\
 & \qquad \delta_{n} \in \{0,1\},  \forall n \in N,\\
 & \qquad \beta_n \in \{0,1\},   \forall n \in N, \label{c5}
\end{flalign}
\end{subequations}
where $\textbf{P}_K=[p_1,\dots,P_K]$ represents transmit power of BS for $K$ UEs, $P_U=[P_1,\dots, P_U]$ represents all transmission power of uplink UEs, $\boldsymbol{\beta}=[\beta_1,\dots\beta_N]$, and $\boldsymbol{\delta}=[\delta_1,\dots,\delta_N]$. 
It can be seen that the formulated optimization problem is a mixed integer nonlinear programming (MINLP) problem. It is highly non-convex, due to the coupling among the variables. In order to tackle this issue, we divide the main optimization problem into two sub-problems. In the first sub-problem, we solve the antenna activation problem. Once the activated PAs are selected, we apply SCA to obtain the power allocation.
\section{Solution Approach}
\subsection{Activated PA Selection}Following \cite{wang2025sum}, we propose an algorithm for antenna selection based on spatial correlation and distance. At the initial stage, we assume that one PA is activated in each waveguide. To minimize path loss, we select the initial PA by calculating the total distance to all users.  Hence, we can calculate the initial PA as follows:
\begin{equation}
    (n^*,k^*)=argmin (d_{n,k}), \forall n \in N, \label{eq19}
\end{equation}
$d_{n,k}=\sum_{k=1}^K||lt_n-u_k||$.
Then, we can activate the PA status as follows.
\begin{equation}
\delta_{n,k} =
\begin{cases}
1, & \text{if } (n,k) = (n^*, k^*), \ \forall n,k, \\[6pt]
0, & \text{otherwise}.
\end{cases}\label{eq20}
\end{equation}
Now, we can calculate the spatial correlation among the users as follows.
\begin{equation}
    \rho =\sum_{k=1}^K \sum_{i=k}^K \frac{|\textbf{h}_k^H\textbf{h}_i|}{\|\textbf{h}_k^H\|\|\textbf{h}_i\|}. \label{eq21}
\end{equation}
We can denote the set of activated PAs as $\Sigma_1$, the set of candidate PAs as $\Sigma_2$, and $\Psi$ is the set of all PAs, where $\Sigma_1 \cup \Sigma_2 = \Psi$. Once the initially activated PAs are selected, they should be removed from $\Sigma_2$ and added to $\Sigma_1$. For the remaining PA selection, we utilize Algorithm 1, which is based on the distance from each user to each PA and the spatial correlation among users to reduce interference. In particular, we repeat the steps until the set of candidate PA $\Sigma_2 = \emptyset$. We measure the distance and spatial correlation based on \eqref{eq19}, \eqref{eq20}, and \eqref{eq21}. The procedure for the proposed method for downlink PA selection is given in Algorithm 1. It should be noted that the same approach is applied to uplink PA selection; details are omitted for brevity. 
\begin{algorithm}[!t]
\caption{PA activation algorithm}\label{alg:two}
\KwData{$\Sigma_1=\emptyset$, $\Sigma_2=\Psi$, $h_{n,k}$, $h_{n,u}$}
\KwResult{The set of $\Sigma_1$}
Adjust $\Sigma_1 \gets \Sigma_1 \cup \eqref{eq19}$\;
Adjust $\Sigma_2 \gets \Sigma_2 \setminus \eqref{eq19}$\;
Calculate $\eqref{eq20}$\;
\While{$\Sigma_2 \neq 0$}{
Calculate $\eqref{eq19}$\;
Adjust $\Sigma_2 \gets \Sigma_2 \setminus (n^*,k^*)$\;
Apply $\eqref{eq20}$ to calculate $\delta_{n,k}$\;
Update $\textbf{h}_k^*$ based on $\eqref{e3}$\;
Calculate $\rho^*$ based on $\eqref{eq20}$\;
  \If{$\rho^* <  \rho$}{
    $\Sigma_1 \gets \Sigma_1 \cup (n^*,k^*)$\;  
  }
  }
\end{algorithm}
\subsection{Power Allocation}
Once activated PAs are selected, the proposed optimization problem becomes a general non-convex problem that can be solved using the conventional SCA-based approach. That being said, problem $\mathcal{P}_1$ can be modeled as follows:
\allowdisplaybreaks
\begin{subequations}
\label{prob:P1}
\begin{flalign}
\centering
 &\mathcal{P}_2: \max_{\substack{\textbf{P}_K, \, \textbf{P}_U}} \quad \quad \sum_{k \in \mathcal{K}} R_{k}+\sum_{u \in \mathcal{U}} R_u,\:\label{c14} \\
 &\text{s.t.} \quad \sum_{k=1}^Kp_k \le P_t,\label{c11}\\
 & \qquad P_{u} \le P_u^{max}, P_{u} \ge 0, \forall u \in \mathcal{U},\label{c12}\\
 & \qquad R_k \ge R_{th,k},  R_u \ge R_{th,u}, \forall k \in \mathcal{K}, \forall u \in \mathcal{U}, \label{c13}
\end{flalign}
\end{subequations}
We see that $\mathcal{P}_2$ is non-convex due to \eqref{c14} and \eqref{c13}. First, we introduce auxiliary variables $\alpha_k$, $\alpha_u$, $\gamma_k$, $\gamma_u$ and replace them in the objective function; hence we get,
\begin{align}
    &\mathcal{P}_2: \max_{\substack{\textbf{P}_K, \, \textbf{P}_U, \boldsymbol{\alpha}_K, \boldsymbol{\alpha}_U, \boldsymbol{\gamma}_K, \boldsymbol{\gamma}_U}} \quad \quad \sum_{k \in \mathcal{K}} \gamma_k + \sum_{u \in \mathcal{U}} \gamma_u,\:\label{c15} \\
    &  \quad \quad \log_2(1+\alpha_k) \ge \gamma_k,\quad \quad \log_2(1+\alpha_u) \ge \gamma_u,\label{beta2}
\end{align}
\begin{equation}
 \frac{|{h}_k|^2 \frac{p_k}{\sum_{n=1}^N\delta_n}}{\sum_{k', k' \neq k}^K|{h}_k|^2\frac{p_k'}{\sum_{n=1}^N\delta_n}+\sum_{u \in U}|h_{u,k}|^2P_u+\sigma_k^2} \ge \alpha_k, \label{c15}
 \end{equation}
 \begin{table*}[!t]
 \begin{equation}
      \frac{\frac{P_u}{\sum_{n=1}^N\beta_n}|{h}_u|^2}{\sum_{u', u' \neq u}^U\frac{P_{u'}}{\sum_{n=1}^N\beta_n}|{h}_{u'}|^2+\frac{1}{\sum_{n=1}^N {\beta_n}}|\hat{\textbf{g}}_r\textbf{H}[TR]\left(\sum_{k \in K}p_k\right)\hat{\textbf{g}}_t^H|^2+\sigma^2/\sum_{n=1}^N\beta_n} \ge \alpha_u, \label{c16}
      \end{equation}
      \end{table*}
where $\boldsymbol{\alpha}_K=[\alpha_k,\dots,\alpha_K]$, $\boldsymbol{\alpha}_U=[\alpha_u,\dots,\alpha_U], \boldsymbol{\gamma}_K=[\gamma_k,\dots,\gamma_K], \boldsymbol{\gamma}_U=[\gamma_u, \dots, \gamma_U], \forall u \in \mathcal{U}, \forall k \in \mathcal{K}$. 
Now it can be seen that \eqref{c15} and \eqref{c16} are still non-convex. First, we introduce two other auxiliary variables such as $\omega$ and $\kappa$ and \eqref{c15} and \eqref{c16} which can be transformed as follows.
\begin{equation}
   \frac{|{h}_k|^2\frac{p_k}{\sum_{n=1}^N\delta_n}} {\omega_k} \ge \alpha_k, \frac{\frac{P_u}{\sum_{n=1}^N\beta_n}|{h}_u|^2}{\kappa_u} \ge \alpha_u,
 \label{c17}
\end{equation}
\begin{equation}
  \omega_k \ge  \sum_{k', k' \neq k}^K|{h}_k|^2\frac{p_k'}{\sum_{n=1}^N\delta_n}+\sum_{u \in U}|h_{u,k}|^2P_u+\sigma_k^2 ,
    \label{c18}
\end{equation}
\begin{table*}[!t]
\begin{equation}
\kappa_u \ge \sum_{u', u' \neq u}^U\frac{P_{u'}}{\sum_{n=1}^N\beta_n}|{h}_{u'}|^2+\frac{1}{\sum_{n=1}^N {\beta_n}}|\hat{\textbf{g}}_r\textbf{H}[TR]\left(\sum_{k \in K}p_k\right)\hat{\textbf{g}}_t^H|^2+
    \frac{\sigma^2}{\sum_{n=1}^N\beta_n}.\label{c20}
\end{equation}\end{table*}
We can see that non-convexity still exists in \eqref{c17}. Following \cite{7946258}, arithmetic and geometric means (AGM) inequality in for any non-negative variables $x,y$ and $z$ and if $xy \le z$, then we have $2xy \le (ax)^2+(\frac{y}{a})^2 \le 2z$ and this inequality only holds if and only if $a=\sqrt{y/x}$. Following this, we can transform \eqref{c17} as follows, 
\begin{equation}
    |{h}_k|^2 \frac{p_k}{\sum_{n=1}^N\delta_n} \ge \alpha_k \omega_k, \label{c40}
\end{equation}
\begin{equation}
  \frac{2|{h}_k|^2}{\sum_{n=1}^N\delta_n} p_k\ge (\alpha_k \zeta_k)^2+(\omega_k/ \zeta_k)^2, \label{c41}
\end{equation}
\begin{equation}
    \frac{P_u}{\sum_{n=1}^N\beta_n}|{h}_u|^2 \ge \alpha_u \kappa_u, \label{c42}
\end{equation}
\begin{equation}
    \frac{2P_u}{\sum_{n=1}^N\beta_n}|{h}_u|^2 \ge (\alpha_u \nu_u)^2+(\kappa_u/ \nu_u)^2, \label{c43}
\end{equation}
where $\zeta_k=\sqrt{\omega_k/\alpha_k}$, $\nu_u=\sqrt{\kappa_u/\alpha_u}$. 
Now using \eqref{beta2}, we can transform \eqref{c13} as follows,
\begin{equation}
    \gamma_k \ge R_{th,k},   \gamma_u \ge R_{th,u}. \label{c45}
\end{equation}
\begin{algorithm}
\caption{Proposed SCA algorithm}\label{alg:two}
\KwData{Initial feasible solutions from the initialization process, $n=0$, tolerance $\epsilon$}
\KwResult{$\textbf{P}_K, \textbf{P}_U, \boldsymbol{\alpha}_K,\boldsymbol{\alpha}_U, \boldsymbol{\gamma}_K, \boldsymbol{\gamma}_U$}
\While{true}{
$n=n+1$\;
Solve $P_3$ using initial feasible solution of $\zeta_k^{n-1}$ and $\nu_u^{n-1}$\;
Denote optimal objective as $\Lambda^n$\;
Update $\zeta_k^{n+1}\gets\zeta_k^{n}$, $\nu_u^{n+1}\gets\nu^n$\;
 \If{$|\Lambda^n -\Lambda^{n-1}| < \epsilon$}{
   break\;  
  }
  }
\end{algorithm}
\begin{figure*}[!t]
\centering
\hspace{-2cm}{\centering\includegraphics[width=0.36\textwidth]{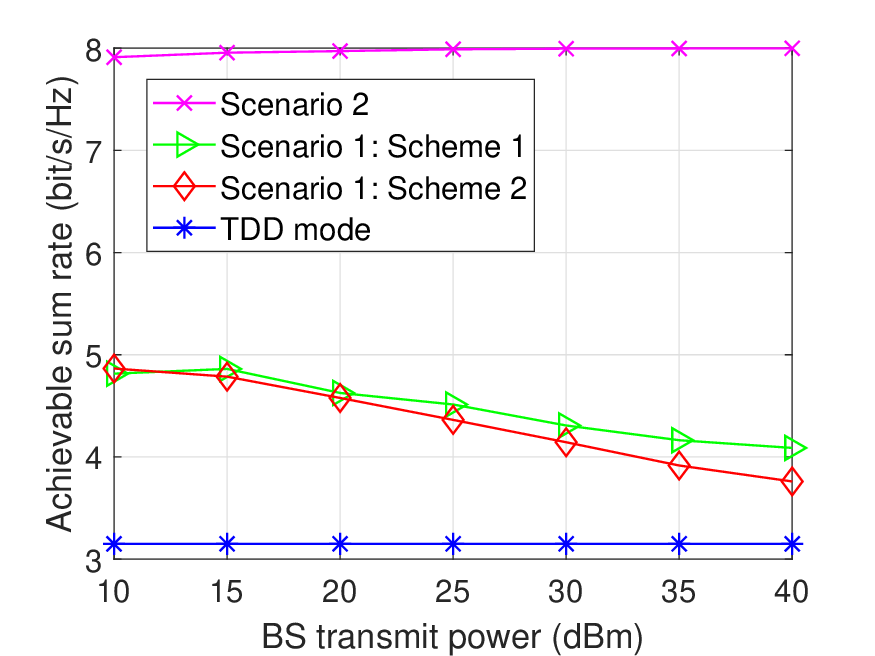}}
\hspace{-0.5cm}{\centering\includegraphics[width=0.36\textwidth]{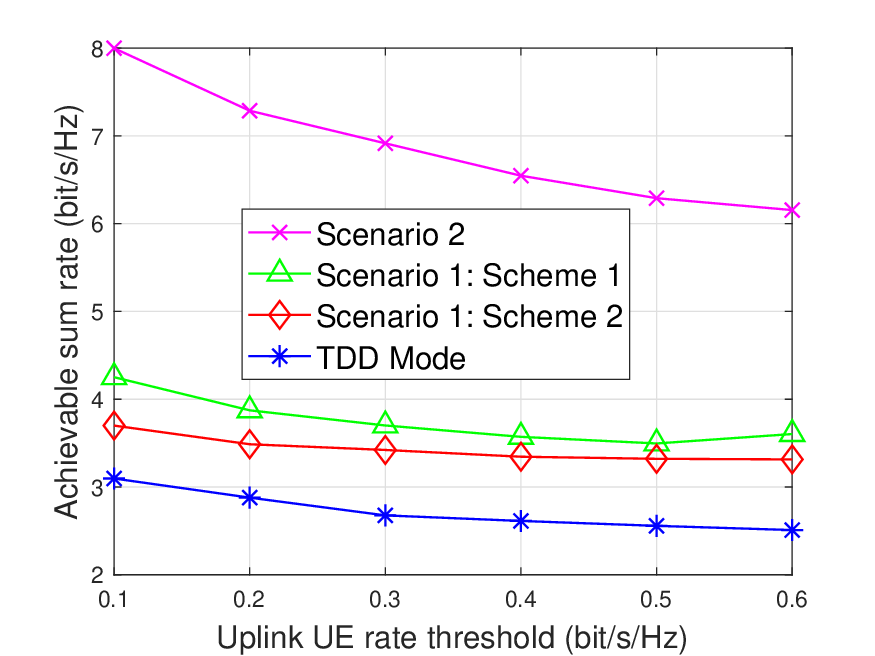}}  \hspace{-0.5cm}
{\centering\includegraphics[width=0.36\textwidth]{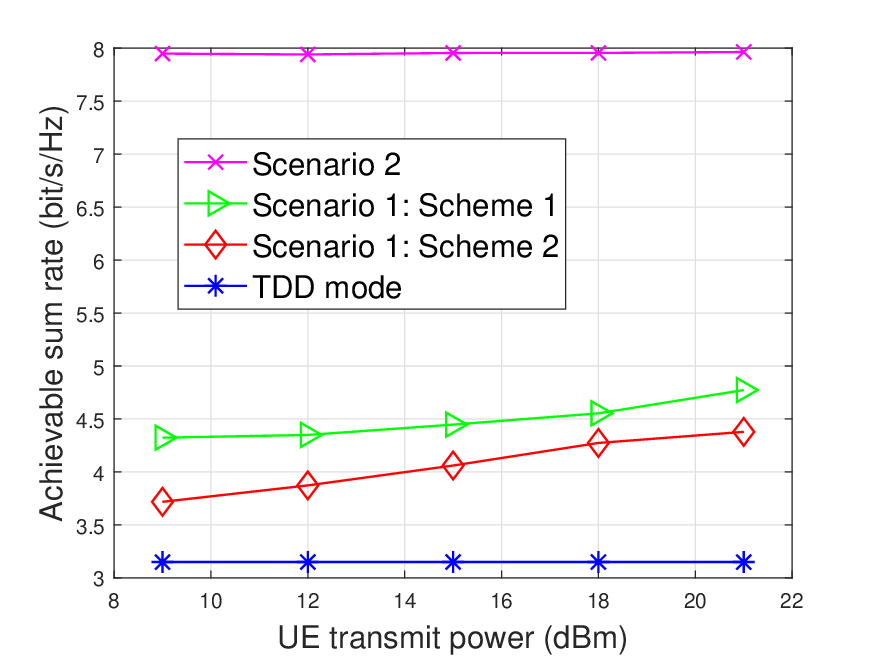}}\hspace{-2cm}
\caption{$R_{th,u}, R_{th,k} =0.1$ bps/Hz, $P_t=40$ dBm, $P_u^{max}=15$ dBm (a) BS transmit power vs total sum rate. (b) Uplink UE rate threshold vs total sum rate. (c) Uplink UE transmit power vs total sum rate.}
\label{fig3}
\vspace{-.2 in}
\end{figure*}
Finally, we can reformulate our problem $\mathcal{P}_2$ as follows,
\allowdisplaybreaks
\begin{subequations}
\label{prob:P3}
\begin{flalign}
\centering
& \mathcal{P}_3: 
\max_{\substack{\textbf{P}_K, \textbf{P}_U, \boldsymbol{\alpha}_K, \boldsymbol{\alpha}_U, \boldsymbol{\gamma}_K, \boldsymbol{\gamma}_U}} 
\quad \sum_{k \in \mathcal{K}} \gamma_k + \sum_{u \in \mathcal{U}} \gamma_u \label{c31} \notag\\
& \text{s.t.} \quad \eqref{c11}, \eqref{c12}, \eqref{beta2}, \eqref{c18}, \eqref{c20}, \eqref{c41}, \eqref{c43}, \eqref{c45}, \notag\\
& \qquad \alpha_k, \alpha_u, \gamma_k, \gamma_u \ge 0, \quad \forall k \in \mathcal{K}, \forall u \in \mathcal{U}, \\
& \qquad \omega_k, \kappa_u > 0, \quad  \forall k \in \mathcal{K}, \forall u \in \mathcal{U}.
\end{flalign}
\end{subequations}
It can be seen that $\mathcal{P}_3$ is convex and can be solved directly with a convex optimization toolbox such as CVX or YALIMP. The problem $\mathcal{P}_3$ is solved iteratively, and the proposed SCA-based scheme is given in Algorithm 2. 
\subsubsection{Initialization process}: We initialize BS and user transmission power randomly within the power budget. For the initialization of $\zeta_k$ and $\nu_u$, we need to initialize $\omega_k$, $\kappa_k$, $\alpha_k$ and $\alpha_u$. $\omega_k$ and $\kappa_u$ are initialized with equality condition of \eqref{c18} and \eqref{c20}.
$\alpha_k$ is initialized with equality condition of \eqref{c15} and $\alpha_u$ is initialized with equality condition of \eqref{c16}.
\subsubsection{Convergence \& Complexity}
The convergence of Algorithm 2 is guaranteed by ensuring that the sequence of objective values ($\Lambda$) forms a monotonically convergent series. Specifically, the SCA framework ensures that the objective function produced at each iteration is monotonically non-decreasing. This property arises because the solution obtained by solving problem $\mathcal{P}_3$ at iteration $n-1$ remains feasible for problem $\mathcal{P}_3$ at iteration $n$. Moreover, since the transmit power budgets at both the BS and the UEs are restricted by constraints \eqref{c11} and \eqref{c12}, respectively, the sequence generated through iterative solutions of problem $\mathcal{P}_3$ is bounded, thereby guaranteeing the convergence of the proposed SCA-based algorithm.
Regarding computational burden, problem $\mathcal{P}_3$ is formulated as a second-order cone program with complexity on the order of $(S_1^2 S_2)$, where $S_1 = (5+N_t)(K+D)+N_t+2$ denotes the total number of variables, and $S_2 = 7(K+D)+2$ represents the number of constraints. Consequently, the overall complexity of Algorithm 1 is given by $O(N_t^2(K+D)^{3.5}\log_2(1/\epsilon))$.
\section{Simulation Results \& Discussions}
In this section, we validate our proposed scheme through numerical analysis. We assume an indoor area of  $D_x \times D_y = 20 \times 1$ m$^2$ where the UEs are randomly distributed and strong LoS exist between UEs and PASS.
The simulation parameters for this work are as follows, $N=10$, $f_c=28$ GHz, $\eta_{eff}=1.4$, $\lambda=0.01$, $\sigma^2=-90$ dBm, $K,U=2$, $h=3$m, $\lambda_x=3$. The distance between two waveguides is $5m$. The pre-configured x-axis position of PAs in transmitting and receiving waveguides are given as follows: $[1,3,5,7,9,12,15,17,19,20]$ and $[0,2,4,6,8,10,11,13,14,18]$. In addition, we assume an equal power distribution model \cite{wang2025modeling}, where the available transmit power at each active antenna is uniformly distributed across all signals. Monte Carlo simulation is performed by averaging over 100 channel realizations. 
We compare our proposed framework with the following baselines.
\begin{itemize}
\item\textbf{Scenario 1: Scheme 1}:  This represents scenario 1, where both waveguides follow the antenna activation policy.
\item\textbf{Scenario 1: Scheme 2}: This is another variation of scenario 1, which is achieved through assuming $\delta_n=1$ and $\beta_n=1$, $\forall n \in N$.  
\item\textbf{Scenario 2}: This represents scenario 2, where no inter-waveguide interference has been considered in this scenario; only interference from uplink UE to downlink UE is considered.
\item \textbf{TDD mode}: we assume that the communication occurs in two time slots. In the first time slot, the downlink communication occurs, and in the second time slot, uplink communication takes place. Hence, the bandwidth for each time slot is halved. However, as the two communications occur in different time slots, there is no inter-waveguide or inter-user interference. 
\end{itemize}
In Fig. \ref{fig3}(a), we investigate the effect of varying the BS transmit power on the achievable sum rate. It can be observed that, as the BS transmit power increases, the sum rate of Scenario 1 (Scheme 1 and Scheme 2) begins to decline. This behavior is primarily attributed to the rise in inter-waveguide interference affecting the uplink UEs. Although a higher BS transmit power provides more downlink transmission capability, it simultaneously amplifies the interference experienced by the uplink UEs, ultimately degrading their achievable rates and reducing the overall system sum rate.

However, the PA selection policy demonstrates a higher ability to mitigate this degradation, as it confines inter-waveguide interference to only a subset of PAs, thereby alleviating the overall interference impact. In addition, it is important to note that all schemes suffer from inter-user interference since no successive interference cancellation (SIC) mechanism is employed. Consequently, both the TDD scheme and Scenario 2 exhibit limited improvement with increasing BS transmit power, as they must balance the trade-off between growing inter-user interference and achievable rate gains.
It should be noted that this effect also influences the Scenario 1-based schemes, which account for the relatively modest increase in their achievable sum rate at the beginning as the BS transmit power rises but decreases after due to the inter-waveguide interferences.

In Fig. \ref{fig3}(b), we examine the effect of varying the uplink UE rate thresholds on the achievable sum rate performance. It can be clearly observed that as the rate threshold increases, the total achievable sum rate of all considered schemes gradually decreases. This trend can be explained by the inherent trade-off between satisfying individual UE rate constraints and maximizing the overall network throughput.
Specifically, when the uplink rate threshold becomes more stringent, each UE must allocate a larger portion of its limited transmit power to ensure that its minimum rate requirement is met. As a result, less power remains available for improving the overall sum rate. This effect becomes more pronounced in scenarios where UEs operate under tight power budgets, as the system’s flexibility in power allocation diminishes. Consequently, the network prioritizes meeting individual rate demands at the expense of sum-rate maximization.

In Fig. \ref{fig3}(c), we present the total achievable sum rate as a function of the uplink UEs’ transmit power. As observed, increasing the UE transmit power leads to an improvement in the sum rate for Scenario 1 (Scheme 1 and Scheme 2). This is because higher transmit power allows the UEs to achieve stronger uplink signals at the BS, thereby enhancing the overall network throughput. In contrast, the Scenario 2 and TDD schemes show negligible improvement with increasing UE transmit power. This behavior can be attributed to the absence of an SIC mechanism, which prevents effective suppression of inter-user interference. As the transmit power rises, the interference among UEs also intensifies, offsetting the potential rate gains from higher transmission levels. Consequently, these schemes reach a practical upper limit in their achievable sum rate, constrained by the balance between maximizing throughput and controlling interference. It is worth noting that this effect also applies to the Scenario 1 schemes, which explains why the rate improvement with increasing UE transmit power is not very pronounced.

In summary, the proposed PA-based communication scheme achieves approximately 90–100\% higher performance compared to its PA-based TDD counterpart when inter-waveguide interference is absent. However, when inter-waveguide interference is taken into account, the performance gain of the PA-based approach is reduced to around 60\%, highlighting the significant impact of interference on overall system efficiency. Meanwhile, it can be observed that the PA selection-based policy achieves a higher sum rate than when we activate all PAs. This is because when a subset of PAs is used, less interference is captured in the receiving waveguide. 
\section{Conclusion}
In this paper, we propose a novel framework of PASS-assisted joint uplink and downlink communication. We jointly optimize antenna activation factor, BS, and UEs transmit power with the objective of maximizing the sum rate. Simulation results demonstrate that our proposed framework can obtain around 60-90\% performance gains over its TDD counterpart. As future work, we plan to enhance this model by incorporating a proportional power model for PASS, accounting for imperfect CSI, and extending it to a multi-waveguide setup. Furthermore, we will investigate the integration of advanced multiple access schemes such as RSMA to develop effective interference management strategies. 
\bibliographystyle{IEEEtran}
\bibliography{references}
\vfill
\end{document}